\documentclass[prl,aps,showpacs,twocolumn,superscriptaddress,amsmath,amssymb]{revtex4}

\usepackage{multirow}
\usepackage{graphicx,dcolumn,bm}
\usepackage{mathrsfs}
\usepackage[usenames,dvipsnames]{color}

\newcommand{\average}[1]{\langle #1 \rangle}

\newcommand{\fr}{\mathbf{r}}
\newcommand{\fR}{\mathbf{R}}

\newcommand{\E}[1]{\hat{E}^{( #1 )}}

\newcommand{\g}[1]{g^{(#1)}}
\newcommand{\ha}{\hat{a}}

\newcommand{\ket}[1]{|#1\rangle}
\newcommand{\bra}[1]{\langle #1|}

\newcommand{\aperture}{\mathcal{A}}

\begin{document}

\title{Super-resolving multi-photon interferences with independent light sources}

\author{S. Oppel}
\affiliation{Institut f\"ur Optik, Information und Photonik, Universit\"at Erlangen-N\"urnberg, 91058 Erlangen, Germany}
\affiliation{Erlangen Graduate School in Advanced Optical Technologies (SAOT), Universit\"at Erlangen-N\"urnberg, 91052 Erlangen, Germany}

\author{T. B\"uttner}
\affiliation{Institut f\"ur Optik, Information und Photonik, Universit\"at Erlangen-N\"urnberg, 91058 Erlangen, Germany}

\author{P. Kok}
\affiliation{Department of Physics and Astronomy, University of Sheffield, Sheffield S3 7RH, United Kingdom}

\author{J. von Zanthier}
\email{Joachim.vonZanthier@physik.uni-erlangen.de}
\affiliation{Institut f\"ur Optik, Information und Photonik, Universit\"at Erlangen-N\"urnberg, 91058 Erlangen, Germany}
\affiliation{Erlangen Graduate School in Advanced Optical Technologies (SAOT), Universit\"at Erlangen-N\"urnberg, 91052 Erlangen, Germany}

\date{\today}

\begin{abstract}
We propose to use multi-photon interferences from statistically independent light sources in combination with linear optical detection techniques to enhance the resolution in imaging. Experimental results with up to five independent thermal light sources confirm this approach to improve the spatial resolution. Since no involved quantum state preparation or detection is required the experiment can be considered an extension of the Hanbury Brown and Twiss experiment for spatial intensity correlations of order $N>2$.
\end{abstract}


\maketitle

Multi-photon interferences with indistinguishable photons from statistically independent light sources are at the focus of current research owing to their potential in quantum information processing \cite{KLM01,Kok07}, creating remote entanglement \cite{Monroe07,Monroe09}, and metrology \cite{Shih10,Shapiro10,Giovannetti11}. The paradigmatic states for multi-photon interference are the highly entangled {\sc noon} states \cite{Dowling2000}, which can be used to achieve increased resolution in spectroscopy, lithography, and  interferometry \cite{Leibfried04,Dowling2000,Shih01,Steinberg04,Walther04}. However, multi-photon interferences from statistically independent emitters---either non-classical or classical---can also lead to enhanced resolution in metrology and imaging \cite{Shih01,Hanbury56,Scully04,Thiel07}. So far, such interferences have been observed with maximally two independent emitters \cite{Shih01,Hanbury74,Zeilinger06,Grangier06,Monroe07a,Yamamoto09,Salomon10,Sandoghdar10,Chekhova2008,Agafonov09,Shih2010}.  Here, we propose to use multi-photon interferences from independent non-classical or classical sources  to obtain spatial interference patterns equivalent to those of {\sc noon} states. Experimental results with up to five independent thermal light sources confirm this approach to enhance the spatial resolution in imaging.

In case of {\sc noon} states the $N$-photon interference pattern can be written as \cite{Dowling2000}
\begin{align}
\label{eq:sr}
  I_N(\mathbf{x}) \propto \frac{1}{2} \left[1 + V_N \cos(N\, \mathbf{k} \, \mathbf{x})\right]\, ,
\end{align}
where $N$ is the number of  photons participating in the {\sc noon} state, $V_N$ is the visibility, $\mathbf{k}$ is the difference vector between the wavevectors $\mathbf{k}_1$ and $\mathbf{k}_2$ of the interfering light fields, and $\mathbf{x}$ is the position along the observation screen. An $N$-photon spatial interference pattern as in Eq.~(\ref{eq:sr}) can be used to enhance the resolution in interferometry and imaging. As known from Abbe, an image of an object is formed if the rays contributing to adjacent diffraction orders (e.~g., $0, +1$) in the diffraction plane are captured by the aperture $\aperture$ of the imaging device since then all information of the object is contained in the diffraction pattern via Fourier transform \cite{BornWolf98}. For a grating with $N$ slits and slit spacing $d$ this leads to a minimal resolvable slit separation $d_{\text{min}}={\lambda}/({2 \, {\aperture}})$, with an error $\Delta d_{\text{min}} = {\lambda}/({4 \, \aperture})$ \cite{BornWolf98}.
This limit can be overcome if the slowly oscillating terms in the diffraction pattern of the grating 
$I \propto 1+\frac{2}{N} \sum_{\alpha=1}^{N-1} (N-\alpha) \cos ( \alpha \delta)$ 
are suppressed such that only the modulation at the highest frequency $\cos \left[(N-1) \delta\right]$ prevails, containing already all relevant parameters of the grating ($N$ and $d$).
Based on counting the number of peaks $M$ across $\aperture$ in the {\sc noon}-like interference pattern $1+ V_N \cos \left[(N-1) \delta \right]$ we obtain $2\pi M = 2 \, \aperture(N-1)kd$. From this, assuming a signal to noise ratio such that $\Delta M < 1/2$, we derive the slit separation $d$ and its error $\Delta d$ as
\begin{align}
\label{Abbe}
 d & = \frac{M\lambda}{2\aperture (N-1)} \\
 \Delta d & = \Delta M \left| \frac{\partial M}{\partial d}\right|^{-1} < \frac{\lambda}{4\aperture (N-1)}\, .
\end{align}
According to Eq.~(\ref{Abbe}), for $N -1 > M$ the pattern conveys information about source details that are smaller than the Abbe limit; for $N -1 > M = 1$ the interference pattern is sensitive to source structures below $\lambda/2$.

\begin{figure}[t!]
\centering
\includegraphics[width=0.48\textwidth]{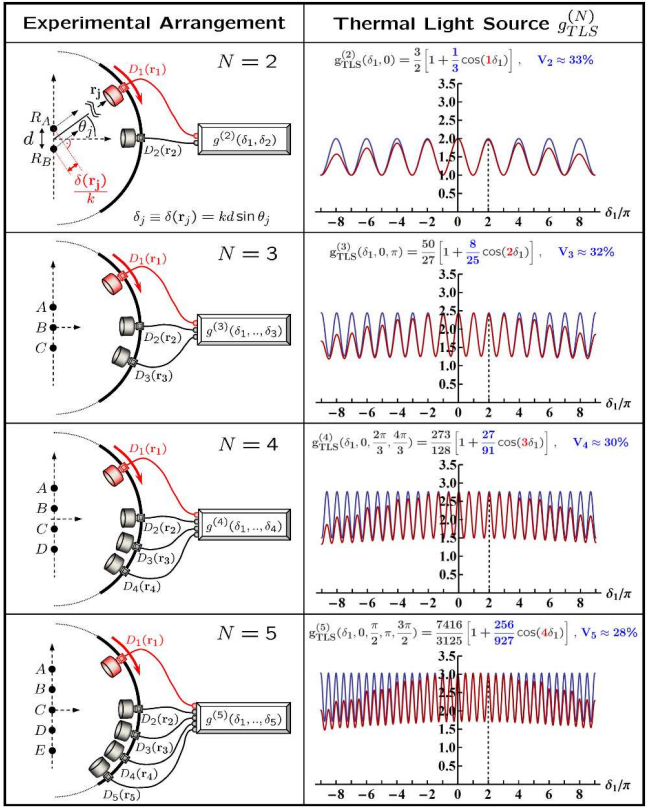}
\caption{Left column: scheme for measuring $\g{N}(\delta_1, \ldots, \delta_N)$ for $N=2, \ldots ,5$ equidistant independent emitters located at $\fR_{\alpha}$ ($\alpha = A, B, \ldots $). $N$ detectors $D_j$ at $\fr_j$ ($j = 1, \ldots, N$) measure $N$ emitted photons in the far field within a joint detection time interval. Right column: Theoretical plots of $\g{N}(\delta_1, \ldots, \delta_N)$ for $N=2, \ldots ,5$ TLS for the indicated fixed detector positions $\delta_j$ ($j=2, \ldots ,N$) as a function of $\delta_1$ for point-like emitters (blue curve) and extended sources (red curve).}
\end{figure}

A super-resolving $N$-photon interference pattern as in Eq.~(\ref{eq:sr}) can be obtained with statistically independent light sources using linear optical detection techniques. Consider $N$ independent emitters at $\fR_{\alpha}$ ($\alpha = A, B, \ldots$) along a chain with equal spacing $d$, and place $N-1$ detectors in a semi-circle in the far field around the sources at specific \textit{magic angles} (see Fig.~1). The emitters are assumed to have identical frequency and polarisation and may be single photon emitters (SPE) or classical thermal light sources (TLS). Moving another detector along the semi-circle and post-selecting on simultaneous single photon detection events in each of the $N$ detectors will produce an interference pattern $I_{N-1}$ as in Eq.~(\ref{eq:sr}), where $\aperture$ is defined with respect to the one detector which is scanned. To see this, we recall that the  $N$-photon interference pattern is proportional to the (normally ordered) $N$-point intensity correlation function
\begin{align}\label{eq:gN}
  g^{(N)}(\fr_1, \ldots, \fr_N) \equiv \frac{\average{:\prod_j \E-(\fr_j) \E+(\fr_j):}}{ \prod_j \average{\E-(\fr_j)\E+(\fr_j) }}\, ,
\end{align}
where $\average{ \cdot}$ denotes the quantum mechanical expectation value, and $\E{+}(\fr)$ and $\E{-}(\fr)$ are the positive and negative frequency parts of the total electric field operator at position ${\bf r}$, respectively. Here, $\E+(\fr_j) \propto \sum_{\alpha} \ha_\alpha\, e^{ikr_{\alpha j}}$, where $\ha_\alpha$ is the annihilation operator of a photon emitted by source $\alpha$ and $r_{\alpha j} = \left| \fR_{\alpha} - \fr_j \right|$ is the distance between the source $\alpha$ and the detector $D_j$. Since the emitters are uncorrelated, the state of the field is given by $\rho = \mathop{\otimes}\limits_{\alpha} \rho_{\alpha}$, where $\rho_\alpha = \sum_n P_\alpha (n) \ket{n}\bra{n}$, with $P_\alpha (n)$ the photon number distribution for the modes originating from source $\alpha$. 

With the detectors at the magic positions, the second- and third-order correlation functions for two and three SPE in terms of phases $\delta_j = k d \sin \theta_j$ reduce to
\begin{align}
\label{eq:qm_2p_tmp}
\g{2}_{SPE}(\delta_1,\delta_2=0) =  \frac{1}{2}\left[1+ \cos(\delta_1)\right] \, ,
\end{align}
\begin{align}\label{eq:qm_3p_si_a}
\g{3}_{SPE}\left(\delta_1,\delta_2=\frac{\pi}{4},\delta_3=\frac{7\pi}{4}\right) = \frac{4}{27}\left[1 + \cos(2\delta_1) \right] \, .
\end{align} 
Here, the visibility in both cases is 100\%; this remains true in case of SPE for any order $N$. In the same way, with the detectors at the magic positions the second- and third-order correlation functions for two and three TLS take the form 
\begin{align}
\label{eq:qm_3p_th_0pi}
	\g{2}_{TLS}(\delta_1,0)  =  \frac{3}{2} \left[1+ \frac{1}{3}\cos(\delta_1)\right] \, ,
\end{align}	
\begin{align}
\label{eq:qm_3p_th_0pi_1}
  \g{3}_{TLS}(\delta_1,0,\pi) = \frac{50}{27} \left[1+ \frac{8}{25}\cos(2\delta_1)\right] \, ,
\end{align}     
displaying a reduced visibility of $V_2 = 1/3$ and $V_3 = 8/25$, respectively, due to the possibility of multiple photons originating from the same TLS. In all cases the super-resolving {\sc noon}-like modulation of $\g{N}(\delta_1)$ as in Eq.~(\ref{eq:sr}) is clearly visible.

Similar results are obtained for higher numbers of SPE and TLS. In the case of TLS, the interference pattern $\smash{\g{N}_{TLS}}$ always reduces to the form $I_{N-1}$ in Eq.~(\ref{eq:sr}) if the $N-1$ fixed detectors are located at the magic positions $\delta_j = 2\pi(j-2)/(N-1)$, $j = 2, \ldots, N$. Note that by changing for different $N$ the angles $\theta_2, \ldots, \theta_N$ of detectors $D_2, \ldots, D_N$ such that the relative phase relation for the magic positions $\delta_j - \delta_{j-1} = 2 \pi/(N-1)$, $j = 3, \ldots, N$, is fulfilled, one can monitor the interference pattern $g^{(N)}_{TLS}(\delta_1)$ until the pure sinusoidal modulation as in Eq.~(\ref{eq:sr}) appears. In this case the number of detectors equals the number of slits, what determines $N$. The  sought-after slit separation $d$ can then be derived from the $\delta_j$ via $d = \lambda/(N-1) (\sin \theta_j - \sin \theta_{j-1})$. With this approach it is possible to determine $N$ and $d$ independently. For $N=2, \ldots ,5$ independent TLS the calculated interference signals $g^{(N)}_{TLS}(\delta_1)$ at the magic positions are displayed in Fig.~1, together with their exact analytical expressions.  

Note that the angular range $\aperture_N = \sin [(\theta_N - \theta_2)/2]$ required by all $N$ detectors is larger than the aperture $\aperture$ needed for detector $D_1$ alone. For a slit separation of $d = \lambda/2$ this is shown in Fig.~2. However, one can see from the figure that $\aperture_N$ always remains smaller than the aperture associated with the classical Abbe limit. Moreover, there is some flexibility in placing the $N-1$ fixed detectors, for example besides or behind the investigated object (assuming $4 \pi$ emission), since the required values for $\delta_j$ ($j = 2,\ldots, N$) are valid modulo $2 \pi$. The red curve representing $\aperture_N = \sin [(\theta_N - \theta_2)/2]$ in Fig.~2b does not take into account this flexibility.

\begin{figure}[b]
 \includegraphics[width=0.48\textwidth]{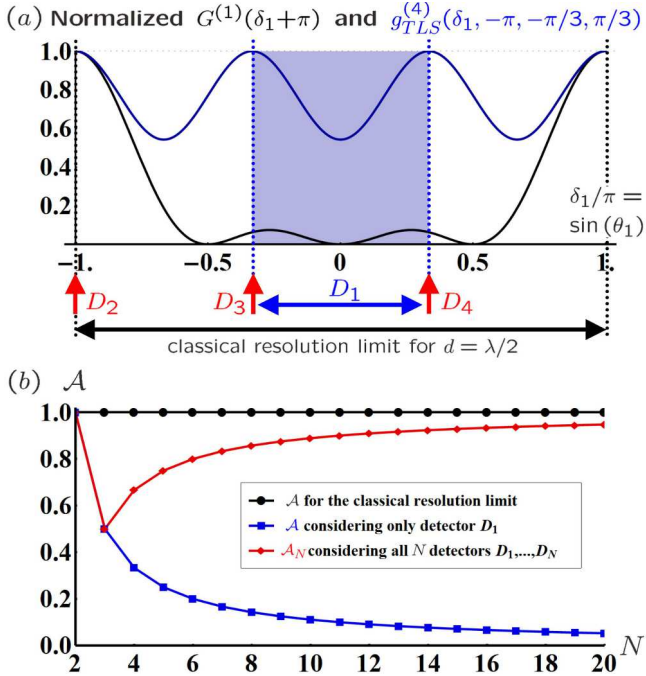}
 \caption{(a) $\smash{\g{4}_{TLS}(\delta_1)}$ for $N=4$ independent TLS (blue curve) with detectors $D_2, D_3$, and $D_4$ at the magic angles $\delta_2=-\pi, \delta_3=-\pi/3,\delta_4=+\pi/3$ (red arrows) and $G^{(1)}(\delta_1+\pi)$ for a coherently illuminated grating with $N=4$ slits (black curve) in case of a source/slit separation $d=\lambda/2$. The angular range required by detector $D_1$ to scan from one to the next principal maximum is indicated for $\smash{\g{4}_{TLS}(\delta_1)}$ by a horizontal blue arrow and for $G^{(1)}(\delta_1+\pi)$ by a horizontal black arrow. The latter is the angular range required by the classical Abbe limit. (b) Numerical apertures required by the classical Abbe limit (black curve), and by the proposed scheme for detector $D_1$ alone (blue curve) and for all $N$ detectors (red curve) to obtain structural information about a grating with $N$ slits and slit separation $d=\lambda/2$.}
 \label{fig:sup1}
\end{figure}

The experimental setup used to measure $\smash{\g{N}_{TLS}}(\delta_1)$ with up to $N=5$ is shown in Fig.~3. To realize the $N$ independent TLS, opaque masks with $N$ identical slits of width $a = 25 \ \mu$m and separation $d = 250 \ \mu$m are illuminated by pseudo-thermal light originating from a linearly polarized Nd:YAG laser at $\lambda=$ 532~nm scattered by a rotating ground glass disk \cite{Rousseau1971}. The large number of time-dependent speckles, produced by the stochastically interfering waves scattered from the granular surface of the disk, act within a given slit as many independent point-like sub-sources equivalent to an ordinary spatial incoherent thermal source. The coherence time of the pseudo-thermal light sources depends on the rotational speed of the disk and was chosen to $\tau_c \approx 100 \ \mu$s. The light from the masks is split by 50/50 non-polarizing beam splitters and collected at a distance $z \approx 1$~m behind the glass disk by $N$ laterally displaceable fiber tips of core diameter 50~$\mu$m, guiding the light to $N$ single photon detectors. The output pulses of the photon detectors are then fed into a coincidence detection circuit. In the experiment the single photon counting rates for $\g{2}_{TLS}$, $\smash{\g{3}_{TLS}}$, $\smash{\g{4}_{TLS}}$, and $\smash{\g{5}_{TLS}}$ correspond to 200~kHz which with joint detection time windows of 50~ns, 410~ns, 410~ns, and 850~ns lead to averaged $N$-fold coincidence rates of 1500~Hz, 1500~Hz, 400~Hz, and 300~Hz, respectively. Note that the $\smash{\g{N}_{TLS}(\delta_1)}$ display the calculated interference signals only if the $N$ photons are measured within their coherence time \cite{Hanbury56}. 

\begin{figure}[t!]
\centering
\includegraphics[width=0.48\textwidth]{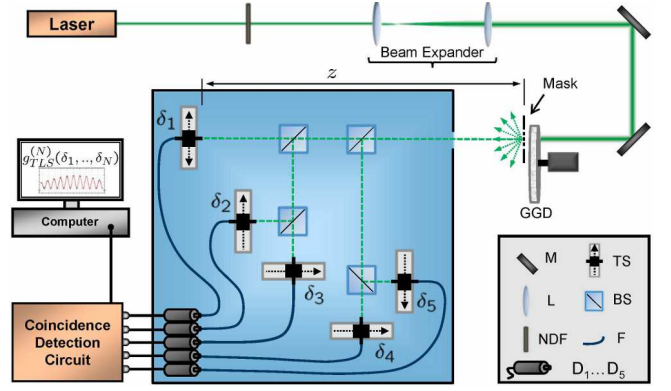}
\caption{Experimental setup for measuring $\smash{\g{N}_{TLS}}(\delta_1)$. For details see text. GGD: ground glass disk, M: mirror, L: lens, NDF: neutral density filter, TS: translation stage with fiber mount, BS: beam splitter, F: multimode fiber, $D_1 \ldots D_5$: photomultiplier modules.}
\end{figure}

The experimental results for $\smash{\g{2}_{TLS}(\delta_1)}, \ldots, \smash{\g{5}_{TLS}(\delta_1)}$ are shown in Fig.~4. The measured curves are in excellent agreement with the theoretical prediction if one takes into account the finite width of the slits [see red solid lines in Fig.~1 and Fig.~4~(b)-(e)]. The small deviations between the experimental results and the theoretical curves for $\smash{\g{4}_{TLS}}$ and $\smash{\g{5}_{TLS}}$ are mostly due to a slight misalignment of the detector positions from the required magic values. The deviations between $\smash{V_N^{(e)}}$ and $V_N$ towards higher $N$ are mainly due to increased dead time effects arising from larger joint detection time windows and higher single photon counting rates at the $N$ detectors. From the figure it can be seen that the measured curves for $\smash{\g{3}_{TLS}(\delta_1)}$, $\smash{\g{4}_{TLS}(\delta_1)}$, and $\smash{\g{5}_{TLS}(\delta_1)}$ display a doubled ($2\delta_1$), tripled ($3\delta_1$), and quadrupled ($4\delta_1$) modulation frequency with respect to $\smash{\g{2}_{SPE}(\delta_1,0)}$ and $\smash{\g{2}_{TLS}(\delta_1,0)}$ [see Eqs.~(\ref{eq:qm_2p_tmp})~and~(\ref{eq:qm_3p_th_0pi})]. This means that for a given aperture $\aperture$ (highlighted in blue in Fig.~4) $\smash{\g{5}_{TLS}(\delta_1)}$ exhibits four times more oscillations than $\smash{\g{2}_{TLS}(\delta_1)}$. According to Eq.~(\ref{Abbe}) this beats the classical Abbe limit for $d$ and $\Delta d$ by a factor of four.

\begin{figure}[t!]
\centering
\includegraphics[width=0.48\textwidth]{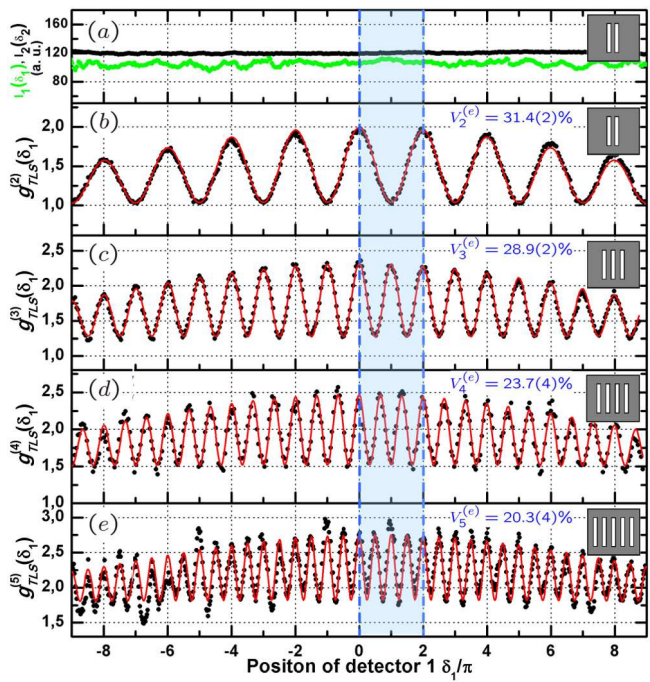}
\caption{ Experimental results: (a) Measurement of average intensities $I_1$ and $I_2$ at detectors $D_1$ and $D_2$ alone (with $D_1$ scanned and $D_2$ kept constant), demonstrating that the pseudo-thermal light used is spatially incoherent in first order of the intensity. (b)-(e) Measurement of $\smash{\g{N}_{TLS}}(\delta_1)$ in case of $N=2, \ldots ,5$ TLS for $\delta_2, \ldots, \delta_N$ at the magic positions. Red curves correspond to a theoretical fit taking into account the finite width of the slits. The only fitting parameters are the slit separation $d$, the slit width $a$, and the visibility $\smash{V^{(e)}_N}$. The experimentally obtained visibilities $\smash{V^{(e)}_N}$ can be compared with the theoretical values $V_N$ in Fig.~1.}
\end{figure}

In conclusion, we experimentally demonstrated spatial multi-photon interference patterns displaying super-resolution with up to five statistically independent light sources using linear optical detection techniques. For $N>2$, these experiments achieve a higher resolution than the classical Abbe limit for imaging the light source. In the case of $N$ SPE, we showed theoretically that the interference pattern obtained is identical to the one generated by {\sc noon} states with $N-1$ photons. The same is true for $N$ TLS, except for a reduced visibility. Our technique neither requires special quantum tailoring of light nor $N$-photon absorbing media, as it relies on single photon detection only. As intensity correlations of order $N>2$ are used to improve the spatial resolution in imaging, it can be regarded a practical extension of the Hanbury Brown and Twiss experiment, one of the fundamental measurement techniques in quantum optics. The natural low light requirements suggest that the technique has potential applications for improved imaging of faint star clusters and in vivo biological samples.

\textit{Acknowledgements} The authors thank I. Harder for the fabrication of the masks and G. S. Agarwal, F. Schmidt-Kaler, A. Ramsay, A. Maser, U. Schilling, and R. Wiegner for critical discussions. The authors gratefully acknowledge funding by the Universit\"atsbund Erlangen-N\"urnberg e.V., the Erlangen Graduate School in Advanced Optical Technologies (SAOT) by the German Research Foundation (DFG) in the framework of the German excellence initiative and the UK Engineering and Physical Sciences Research Council. S.O. thanks the Elite Network of Bavaria for financial support.

\end{document}